\documentclass{elsart}
\usepackage{graphicx}
\usepackage{amssymb}

\begin{document}
\intextsep 30pt
\textfloatsep 30pt
\begin{frontmatter}
\title{Performance of a micro-TPC for a time-resolved neutron 
PSD}
\author[1]{\corauthref{cor1}Kentaro~Miuchi}\ead{miuchi@cr.scphys.kyoto-u.ac.jp},
\author[1]{Hidetoshi~Kubo},
\author[1]{Tsutomu~Nagayoshi},
 \author[1]{Reiko~Orito},
 \author[1]{Atsushi~Takada},
 \author[1]{Atsushi~Takeda},
 \author[1]{Toru~Tanimori},
 \author[1]{Masaru~Ueno}
\address[1]{Cosmic-Ray Group, Department of Physics, Faculty of Science, Kyoto University Kitashirakawa, Sakyo-ku, Kyoto, 606-8502, Japan}
\corauth[cor1]{Corresponding author.\\~~~Tel.$:$+81(0)75-753-3867$\;$fax$:$+81(0)75-753-3799}
\begin{abstract}
We report on the performance of a micro-TPC with 
a micro pixel chamber($\mu$-PIC) readout for a 
time-resolved neutron position-sensitive detector(PSD).
Three-dimensional tracks and the Bragg curves of protons 
with energies of around 1 MeV
were clearly detected by the micro-TPC.
More than 95$\%$ of gamma-rays of 511 keV 
were found to be discriminated by simple analysis. 
Simulation studies showed that 
the total track length of proton and triton emitted from the 
$\rm {}^{3}He$(n,p(573 keV))$\rm {}^{3}H(191 keV)$ reaction
is about 1.2 cm, and that both particles 
have large energy losses ($\rm > 200 keV/cm$) 
in 1 atm Ar+$\rm C_{2}H_{6}(10\%)$+${}^{3}$He($< 1\%$). 
These values suit the current performance of the micro-TPC, and  
we conclude that a time-resolved neutron PSD with 
spatial resolution of sub-millimeters
shall be developed as an application of the micro-TPC.
\end{abstract}

\end{frontmatter}

\newpage
\section{Introduction}
\label{intro}
Micro-TPC, a time projection chamber with a micro pixel chamber ($\mu$-PIC)
 readout was recently developed 
for the detection of charged particles\cite{TPC_PSD:Kubo}. 
The results of a fundamental measurement indicate
that some improvements are still required for detecting  
the minimum ionizing particles (MIPs)\cite{IEEE:Miuchi}.
Nevertheless, the results also indicate that 
low-energy charged particles, which have a large energy loss ($dE/dx$), 
are already detectable by the micro-TPC with dense samplings. 

Neutron position-sensitive detectors (PSDs) 
with a large detection area and 
a capacity for high-rate operation
are indispensable for use at neutron beams 
of the next generation\cite{netron_summary:Eijk}. 
Since the neutron energy is resolved by
a measurement of the time-of-flight(TOF), a 
timing measurement on the order of $\mu$s is strictly required. 
Gaseous neutron detectors filled with $\rm{}^{3}He$ have been keenly 
developed and used because of large cross section
 to thermal neutrons of about 25 meV.
Because the energy losses of both the proton and the triton 
emitted from the 
$\rm {}^{3}He$ (n, p (573 keV)) $\rm {}^{3}H (191 keV)$ reaction
are large, 
we expect to detect the fine tracks of 
both particles by the micro-TPC.
We can thus determine the incident position from the detected tracks 
with a resolution of sub-millimeters.
Since gamma-rays are thought to make a large background when used at 
neutron beams,
discrimination between the neutrons and gamma-rays is also important,
which should be realized by the measured values, such as 
the energy losses ($dE/dx$).

In this paper, we present the tracking performance of the micro-TPC 
with energies of around 1 MeV.
We then report on the discrimination 
between the electron tracks (gamma-ray events) and 
the proton tracks (neutron events). 
Finally, we discuss the development of 
time-resolved thermal neutron ($\sim$ 25 meV) PSD 
as an application of the micro-TPC 
while taking account of the experimental results and simulation studies.

\section{Micro-TPC}
\label{micro-TPC}
\begin{figure}[p]
   \begin{center}
	\includegraphics[width=0.7\linewidth]{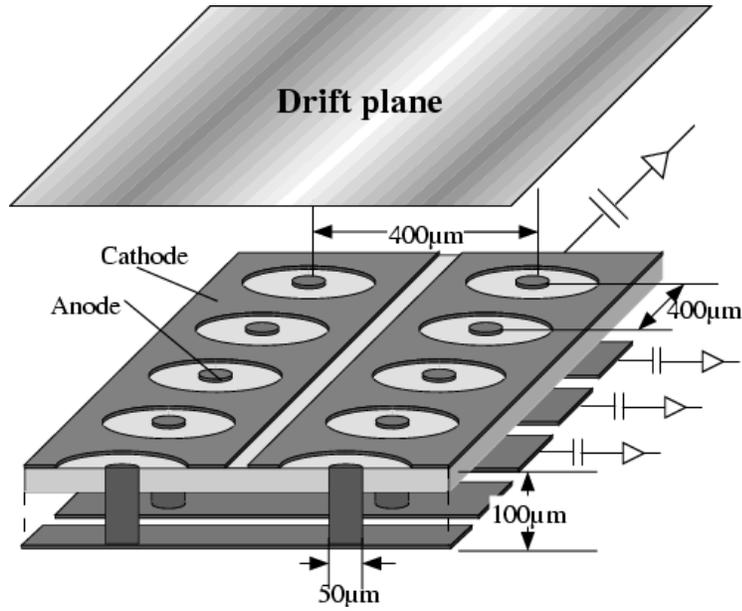}
   \caption{Schematic structure of the $\mu$-PIC. 
Cathode strips are formed on one side of 
the polyimide substrate of 100 $\mu$m thick, while anode strips 
are orthogonally formed on the other side.}
   \label{fig:uPIC}
  \end{center}
   \end{figure}

A micro-TPC, a time projection chamber with a micro pixel chamber($\mu$-PIC\cite{uPIC:Ochi}) readout, was recently developed for the detection of charged particles\cite{TPC_PSD:Kubo,IEEE:Miuchi}. 
$\mu$-PIC is a gaseous two-dimensional PSD 
manufactured by printed circuit board(PCB) technology. 
A schematic of the $\mu$-PIC structure is shown in Fig.~\ref{fig:uPIC}. 
We developed a $\mu$-PIC with a detection area of 10 $\times$ 10 $\rm cm^2$  and a pixel pitch of 400 $\mu$m.  Cathode strips are formed on one side of 
a polyimide substrate (100 $\mu$m thick), while anode strips 
are orthogonally formed on the other side.

The results of a fundamental measurement have already been 
reported in a previous paper\cite{IEEE:Miuchi}. Here, we briefly mention the essential features of the micro-TPC\cite{TPC_PSD:Kubo}. The detection volume of the micro-TPC is $\rm 10 \times 10 \times 8 cm^3$ with a drift field of 0.4kV/cm. The micro-TPC 
can be stably operated at a gas gain of 3000 with Ar-$\rm C_2H_6(10\%)$ flow.
We are optimizing the geometrical structure of the pixels
using a three-dimensional simulator\cite{Maxwell,Garfield}
in order to realize a stable operation with a gas gain of ${10}^{4}$.  
The signal from each strip 
is fed to an amplifier-shaper-discriminator (ASD\cite{ASD:Sasaki}) chip, 
which outputs both an amplified analog signal and a discriminated digital (LVDS) signal.
LVDS signals are in turn read 
by a position encoding module (PEM),  which works at a clock rate of 20 MHz. 
PEM calculates the
two-dimensional incident position 
while taking the anode-cathode coincidence within one clock pulse.
When at least one anode-cathode coincidence is found 
within the maximum drift time (2 $\mu$s) from the 
external trigger, or the ''$t=0$'' time, 
the two-dimensional position and the elapsed time from the trigger 
are recorded.
For the energy measurement, every 32 analog outputs of the cathode ASD chips are summed and digitized by an 8-channel 100 MHz flash ADC (FADC).
In this way, we can realize three-dimensional tracking and spectroscopy of the charged particles with the micro-TPC.

\section{Measurements}
\label{measurements}
\subsection{Proton tracking}
\label{section:tracking}
\begin{figure}[p]
   \begin{center}
\includegraphics[width=0.7\linewidth]{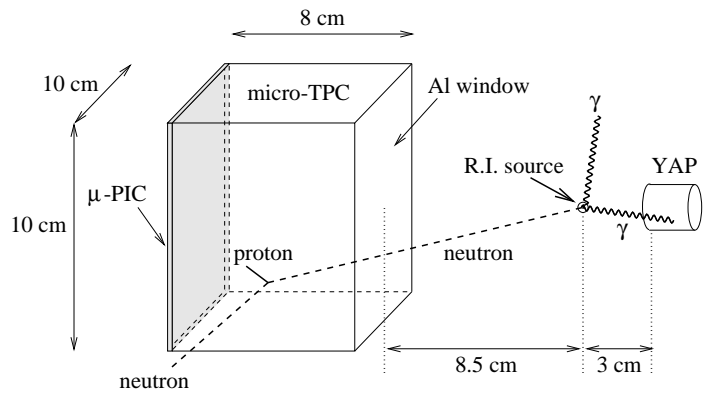}
   \caption{Schematic drawing of the experimental set-up. ${}^{252}$Cf is placed at the ''R.I. source'' position in the neutron-run, while
${}^{22}$Na is used in the $\gamma$-run.}
   \label{fig:setup}
  \end{center}
   \end{figure}

We irradiated the micro-TPC with 
fast neutrons from a radioactive $\rm{}^{252}Cf$($\sim$2MBq) source, 
the neutron energy of which peaks between 500 keV and 1 MeV.
Protons arise from elastic neutron scattering by hydrogen nuclei in
$\rm C_2H_6$. Part of the neutron kinetic energy is transferred to 
the hydrogen nucleus, i.e. the proton. 
The radioactive source was placed 8.5 cm 
from the aluminum window of the micro-TPC, and an Yttrium Aluminum Perovskite 
(YAP) scintillator\cite{Knoll}(1''$\times$ 1'' $\phi$) 
was set 3 cm from the radioactive source 
at the opposite side of the micro-TPC. 
The set-up is shown in Fig. \ref{fig:setup}.
One fission decay of $\rm{}^{252}Cf$ emits 3.8 neutrons and 9.7 photons
on average,  
therefore, the micro-TPC was triggered by 
gamma-rays detected by the YAP scintillator.
The three-dimensional tracking performance for protons 
of around 1 MeV and a gamma-ray background of several hundred keV
were thus measured in this ''n/$\gamma$-run''.

From all of the measured data, we selected data with a  
$length > $1cm and $N_{hit}\geq$4 as the track data.
Here, $length$ is the track length calculated by 
simply connecting the detected points, 
and $N_{hit}$ is the number of detected points.

\begin{figure}[p]
   \begin{center}
\includegraphics[width=0.7\linewidth]{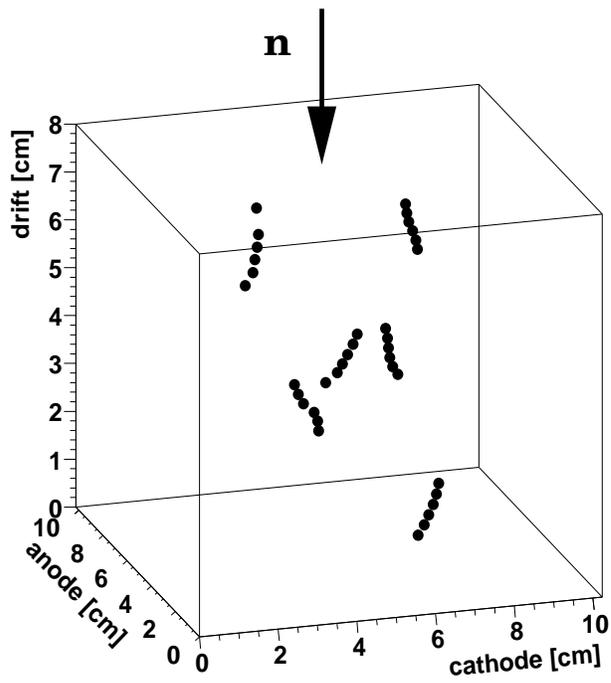}
   \caption{Several three-dimensional proton tracks(500 keV - 1 MeV) detected in the n/$\gamma$-run. }
   \label{n_tracks}
  \end{center}
   \end{figure}

\begin{figure}[p]
   \begin{center}
\includegraphics[width=0.7\linewidth]{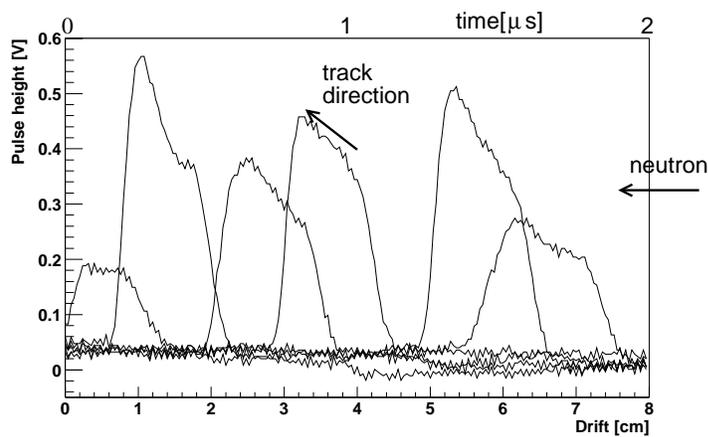}
   \caption{Energy loss of the protons. Each track has its counterpart in Fig. \ref{n_tracks}.
The directions of the tracks are recognized from the shape of the Bragg curves.
}
   \label{Edep_track}
  \end{center}
   \end{figure}

Several proton tracks with energies between 500 keV and 1 MeV 
are shown in Fig. \ref{n_tracks}. 
The hardware threshold level at the ASD chip was 50 keV/cm, which was low enough to detect sub-MeV protons having energy losses larger than 200 keV/cm. Therefore, the detection efficiency was estimated to be almost 100$\%$.
In Fig. \ref{Edep_track}, 
FADC waveforms of the 
proton tracks are shown.
Here, the same events are shown in Fig. \ref{n_tracks} 
and Fig. \ref{Edep_track}.
Since the elapsed time from the trigger represents the 
drift length, these waveforms are 
regarded as the Bragg curves.
The directions of the tracks are obviously 
known from the shape of the Bragg curves.

From measurements with $\rm{}^{252}Cf$, 
the micro-TPC was found to possess sufficient
performance to 
detect the tracks and Bragg curves
of protons with energies of around 1 MeV.

\subsection{Particle discrimination}
Since gamma-rays are thought to make a large background 
when used at neutron beams,
discrimination between the neutrons and gamma-rays is important.
We measured the particle discrimination power 
by irradiating the micro-TPC with  
gamma-rays from a radioactive source of $\rm{}^{22}Na$. 
Annihilated back-to-back gamma-rays of 511 keV were emitted from the source.
The micro-TPC was triggered by one of the gamma-rays 
detected by a YAP scintillator, 
while the other scattered the electrons in the micro-TPC.  
This radioactive source was chosen because the gamma-ray energy is 
close to the $Q$ value (764 keV) of the $\rm {}^{3}He$ (n, p) $\rm {}^{3}H$ reaction. 
Thus, the particle discrimination of the 511 keV gamma-rays was measured in this ''$\gamma$-run''. 
Since the $dE/dx$ of electrons scattered by the gamma-rays were much smaller than those of the  neutrons, analog signals were amplified by the gain amplifier((gain = 8) before being digitized by the FADC in the $\gamma$-run.

\begin{figure}[p]
   \begin{center}
\includegraphics[width=0.7\linewidth]{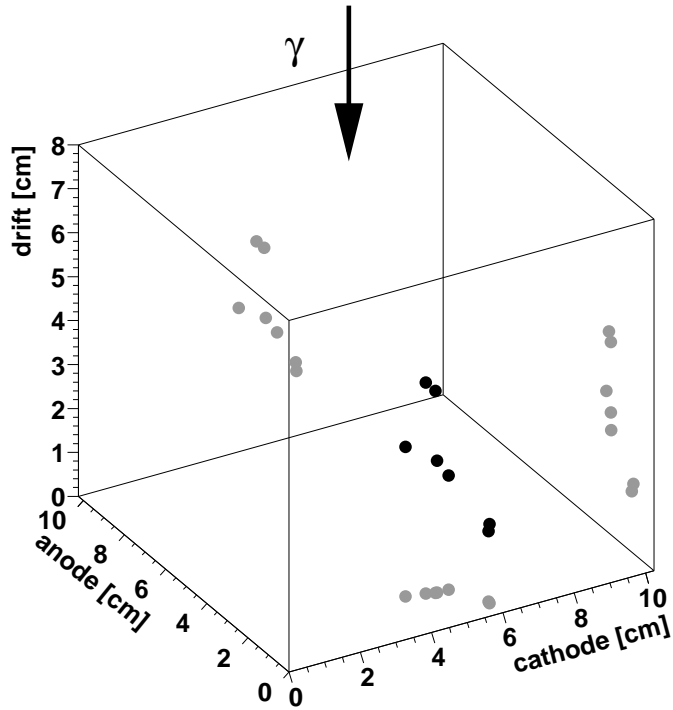}
   \caption{Typical three-dimensional track of an electron detected in the $\gamma$-run. Projections are also shown in the gray points.}
   \label{gamma_tracks}

  \end{center}
   \end{figure}

\begin{figure}[p]
   \begin{center}
\includegraphics[width=0.9\linewidth]{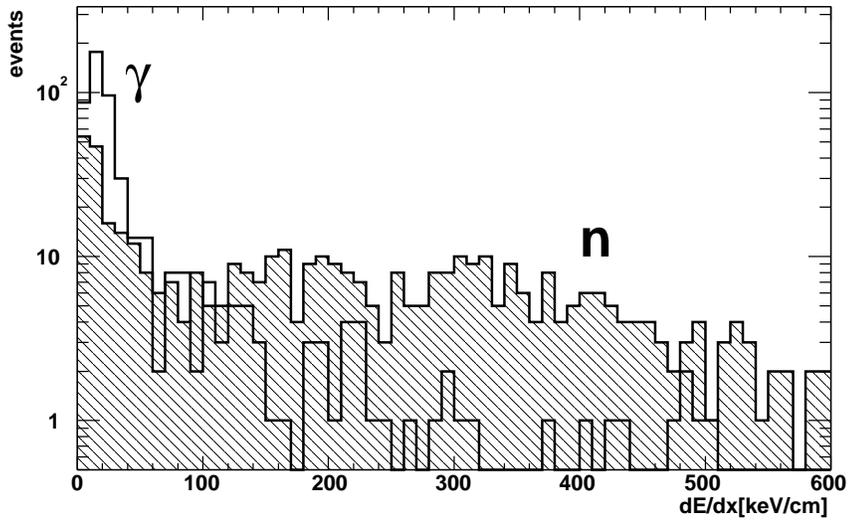}
   \caption{$dE/dx$ distributions of the n/$\gamma$-run(hatched) and $\gamma$-run(non-hatched.) Neutron events are seen in $>$50 keV/cm in the n/$\gamma$-run.}
   \label{dEdx_hist}
  \end{center}
   \end{figure}

\begin{figure}[p]
   \begin{center}
\includegraphics[width=0.9\linewidth]{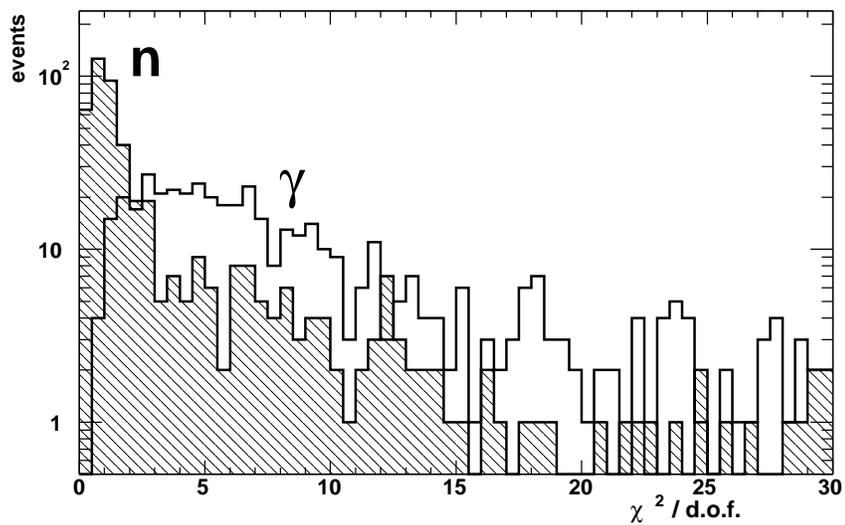}
   \caption{$\chi ^2/$d.o.f. distributions of the n/$\gamma$-run(hatched) and $\gamma$-run(non-hatched.) Neutron events make the peak below 3.}
   \label{X2_hist}
  \end{center}
   \end{figure}

The typical three-dimensional track of the electron (gamma-ray event) is shown in Fig. \ref{gamma_tracks}. 
Comparing the tracks in Fig. \ref{n_tracks} and Fig. \ref{gamma_tracks},
one finds that the proton tracks are more dense and straight than those of the electrons.  Consequently, one can assume that the discrimination of the proton tracks and the electron tracks are realized by the energy loss and the 
fitting results with straight lines.
We defined the energy loss by $dE/dx=E/length$ and $\chi ^2$ by
\begin{equation}
\chi ^2 = \sum_{i=1,2,n-1,n} \frac{\Delta _i}{\sigma},
\end{equation}
where  $E$ is the detected energy, n is the number of detected points, $\Delta_i$ is the distance between the $i$th detected point and the best-fit straight line, and $\sigma$ = 270$\mu$m is the measured three-dimensional spatial resolution of the micro-TPC. The degree of freedom (d.o.f) was three for all of the tracks, since we used the first two and the last two points to calculate $\chi^2$.
The $dE/dx$ distributions of the n/$\gamma$-run and the $\gamma$-run are shown in Fig. \ref{dEdx_hist}.  In the data of the $\gamma$-run, most events are distributed below 50 keV/cm.
On the other hand, neutron events with $dE/dx>$ 50 keV/cm can be seen in the n/$\gamma$-run as well as the gamma-ray peak below 50 keV/cm.
The $\chi ^2/$d.o.f distributions are shown in Fig. \ref{X2_hist}. $\chi^2/$d.o.f distribution of the n/$\gamma$-run peaks below 3, because the proton tracks are fitted with the straight lines very well. We thus define the ''neutron region'' by $dE/dx>$50 keV/cm and $\chi^2/$d.o.f$<$ 3.  
When we consider the selection efficiency of the neutrons,
the lower limit for the energy loss ($dE/dx=$50 keV/cm) is reasonably low 
compared to the energy losses ($\rm > 200 keV/cm$) of the protons and 
tritons from the $\rm {}^{3}He$(n,p)$\rm {}^{3}H$ reaction. 
It is apparent that the $\chi^2$ cut 
has a very high efficiency from the steep peak in 
Fig. \ref{X2_hist}.
Therefore, the selection efficiency for the neutron is thought to be very close to 100$\%$.

\begin{figure}[p]
   \begin{center}
\includegraphics[width=0.9\linewidth]{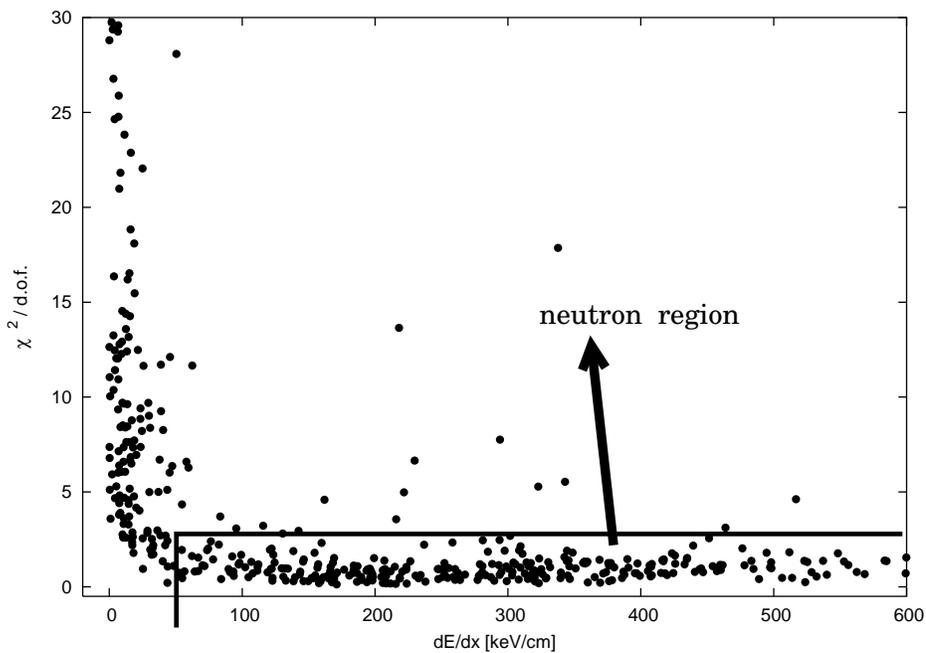}
   \caption{$\chi^2/$d.o.f vs $dE/dx$ plots of the n/$\gamma$-run. The neutron region is superimposed.}
   \label{n_scat}
  \end{center}
   \end{figure}

\begin{figure}[p]
   \begin{center}
\includegraphics[width=0.9\linewidth]{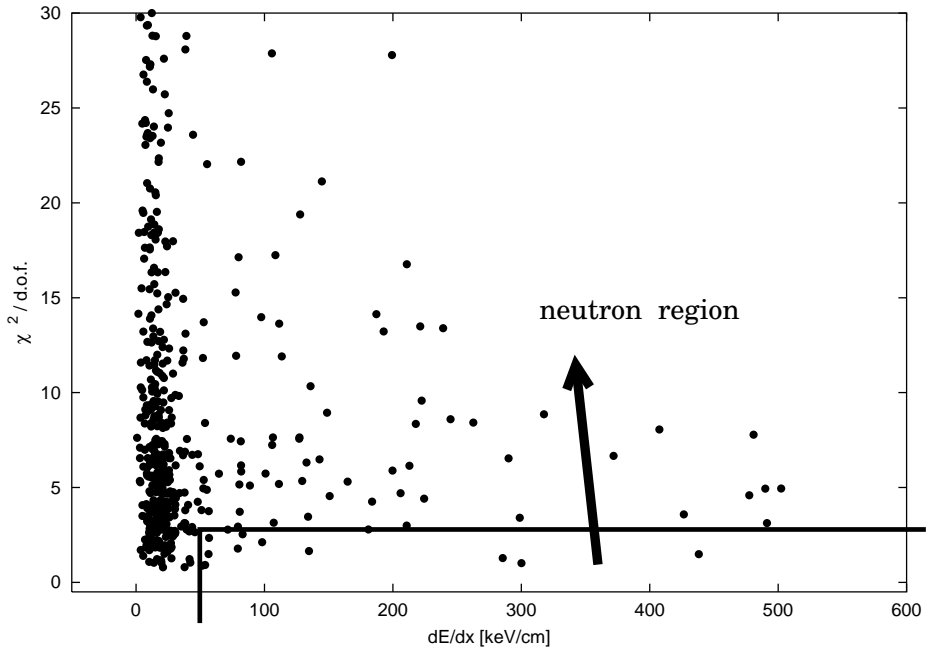}
   \caption{$\chi^2 /$d.o.f. vs $dE/dx$ plots of the $\gamma$-run. Less than 
5$\%$ of the detected events (14 of 500 events) are seen in the superimposed 
neutron region.}
   \label{g_scat}
  \end{center}
   \end{figure}

Fig. \ref{n_scat} shows $\chi ^2/$d.o.f. vs $dE/dx$ plots of the 
n/$\gamma$-run. Neutron events can be seen in the neutron region, while the 
gamma-ray events are seen out of the neutron region. 
The result of the $\gamma$-run is shown in Fig. \ref{g_scat}.  Only less than 5$\%$ of the total events (14 of 500 events) are seen in the neutron region, which indicates that more than 95$\%$ of the gamma-rays are discriminated  by this analysis.

In this measurement with a $\rm{}^{22}Na$ radioactive source, 
more than 95$\%$ of the 511 keV gamma-ray background is known to be 
discriminated by the $dE/dx$ and the $\chi^2$.

\section{Time-resolved neutron PSD with the  micro-TPC} 
\subsection{Time-resolved neutron PSD with the micro-TPC} 
Neutron position-sensitive detectors(PSDs) 
with a large detection area and 
a capacity for high-rate operation
are indispensable for use at neutron beams 
of the next generation\cite{netron_summary:Eijk}. 
Since the neutron energy is resolved by
a measurement of a time-of-flight(TOF), the 
timing measurement on the order of $\mu$s is strictly required. 
Gaseous neutron detectors filled with $\rm{}^{3}He$ have been keenly 
developed and used because of the large cross section
 to thermal neutrons of about 25 meV.
Recently, a CCD-GEM based $\rm{}^{3}He$ detector was 
developed, and the performance was studied\cite{CCD-GEM:Fraga}. 
Nice images of the tracks of the proton and triton 
were obtained, which indicates the potential for the thermal-neutron PSD.
However, making a large-area detector and high-rate operation could 
be problematic for practical use at neutron beams. In addition, the CCD readout is too slow for the TPC; hence, only two-dimensional tracks are achieved in this 
readout system. This feature would deteriorate the quality of the neutron images.

On the other hand, the micro-TPC with a large detection area 
is easily manufactured and high-rate operation up to 7.7 MHz with 
the $\mu$-PIC was 
actually realized\cite{IEEE:Miuchi}. 
Three-dimensional trackings help to determine the incident position
with a spatial resolution of sub-millimeters. 
Fine spatial resolution is strictly required for the 
neutron diffraction imaging,  
because the incident angle is determined by the incident position.
Therefore, the micro-TPC is 
an appropriate detector for time-resolved neutron PSD
with $\rm {}^{3}He$. The principle of the $\rm {}^{3}He$ neutron detector is $\rm {}^{3}He$ (n, p (573 keV)) $\rm {}^{3}H (191 keV)$.
We have already shown that the micro-TPC possess sufficient
performance to detect the tracks and Bragg curves
of the proton emitted from this reaction.
We subsequently studied the 
tracks of both particles by a simulation, and 
evaluated the development of the 
time-resolved thermal neutron ($\sim$ 25 meV) PSD 
as an application of the micro-TPC.

\begin{figure}[p]
   \begin{center}
\includegraphics[width=0.9\linewidth]{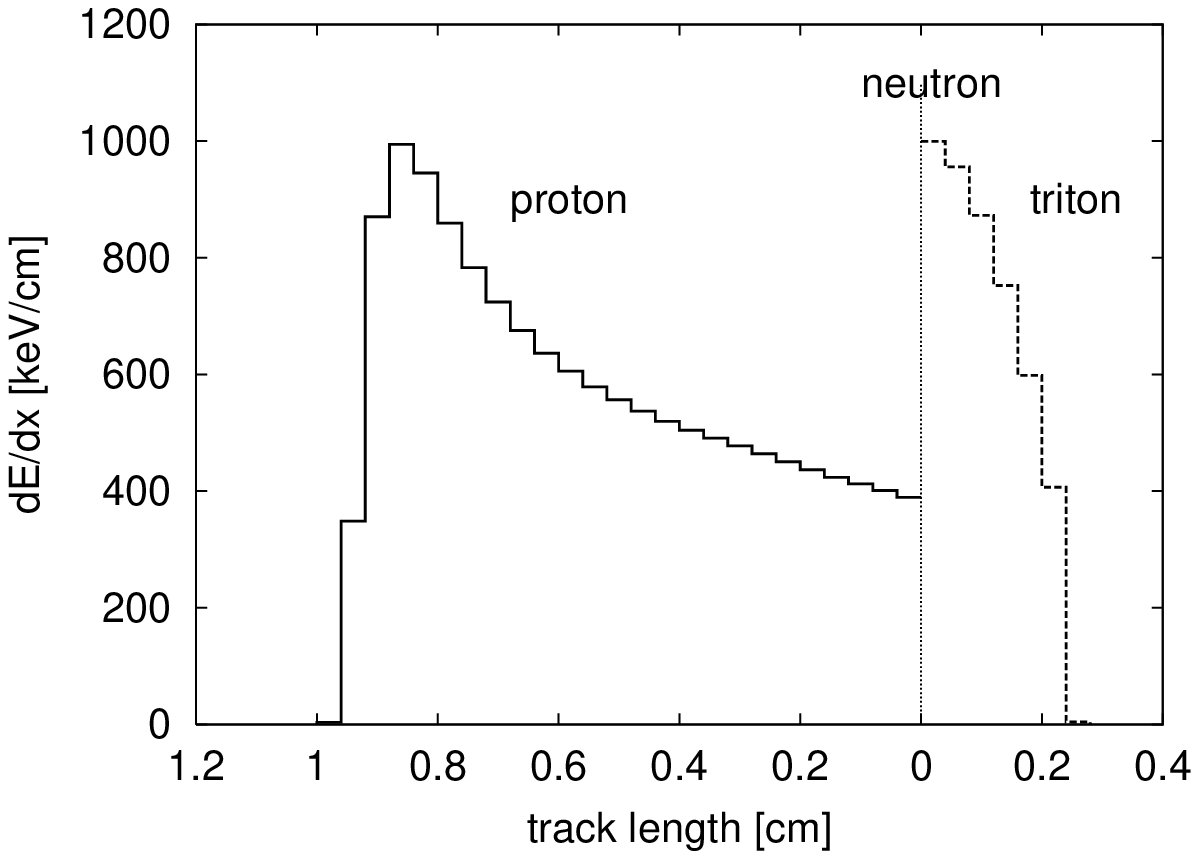}
   \caption{Calculated energy loss along the proton and triton track in 1 atm Ar+$\rm C_{2}H_{6}(10\%)$+${}^{3}$He($< 1\%$). The total track length is about 1.2 cm.}
   \label{pro_tri_track}
  \end{center}
   \end{figure}

We calculated the energy depositions of protons and tritons along the tracks by Geant4 (ver 5.0 patch-01)\cite{geant4}.  A gas mixture of  Ar-$\rm C_2H_6(10\%)$-$\rm{}^{3}He$($<1\%$) at 1 atm was used for the calculation. We did not take account of the 
ionization of ${}^{3}$He for the energy deposition 
because its amount is vary small. 
The result is shown in Fig.  \ref{pro_tri_track}. The track length (1.2cm) and the $dE/dx$ of both particles ($>$ 200 keV/cm) are reasonable for detection by the micro-TPC. The protons and the tritons are easily distinguished 
from the Bragg curves.
As a result, the incident position is determined with a spatial resolution of sub-millimeters.

From a measurement with a ${}^{22}$Na source, 
more than 95$\%$ of the 511 keV gamma-ray background was 
known to be discriminated 
by $dE/dx$ and $\chi^2$.
For practical use, the total energy deposition ($E$) is 
also used to discriminate the low energy gamma-rays that have large $dE/dx$. 
With this total energy cut,  almost complete gamma-ray rejection will 
be realized.
In this way, we reject the gamma-ray background
almost completely, which is another
appropriate feature of the neutron PSD as an application of the micro-TPC.

This neutron PSD is operated with the gas at normal pressure. 
We can thus reduce the materials needed for the high-pressure gas enclosure,
which is useful for a better image of the neutrons.

From simulation studies and the experimental results, we conclude that a time-resolved neutron PSD with 
a spatial resolution of sub-millimeters
shall be developed as an application of the micro-TPC.  

\subsection{Future plans}
We are developing a $\mu$-PIC with a detection area of 30 $\times $ 30 cm${}^{2}$. 
We will soon increase the clock rate of the encoding system from 20 MHz to 100 MHz, because the current spatial resolution is dominated by this clock rate. 
With these improvements, we are quite sure
that we can develop a time-resolved neutron PSD as an aplication 
of the micro-TPC.

The parallax error could be problematic for a non-pressurized 8cm-thick TPC.
One solution to avoid it is to build a curved detector so that one of the 
two parallax errors will not be observed. 
Because the $\mu$-PIC is a thin ($\sim$100$\mu$m) polyimide sheet, 
curved detectors can, in principle, be manufactured.
Another solution is to determine the start timing of the drift 
by detecting the gas scintillation light so that 
the interaction position would 
be determined three-dimensionally with a resolution of sub-millimeters. 
The study on the gas scintillation 
seems to be one of the most important tasks of ours in the 
near future development.
$\rm CF_4$ gas, which is a common gas for the $\rm{}^{3}He$ neutron detector 
because of its large stopping power (for shorter tracks) 
and small Z (for less gamma-ray background), would also be 
useful for this purpose, 
because its scintillation wavelength fits the 
detection by the photomultipliers\cite{ref:GEM_Fraga}. 
Total light yield from the $\rm CF_4$ gas 
by the neutron capture reaction of $\rm {}^{3}He$ 
is estimated to be $O (10^3)$ photons, 
which seems to be enough to trigger the TPC with a 
timing resolution of $\sim$10ns even 
with a photomultiplier coverage of a few $\%$ and 
its quantum efficiency of $\sim 10\%$ at 600nm.
In this way, we think the parallax error could be avoided 
with some more improvements of the micro-TPC.



The results on the tracking performance and the particle discrimination 
indicate a strong possibility the application of the micro-TPC
as a dark matter detector.  Actually, our results are comparable to 
those shown in Ref. \cite{DRIFT_NIM} concerning the points of tracking 
performance and discrimination. 
We are going to study the 
detector response to low-energy nuclear recoils in order to 
estimate the feasibility for a dark matter detector.

%

\section{Conclusions}
\label{conclusions}
Three-dimensional tracks and the Bragg curves of the protons 
with energies of around 1 MeV
were clearly detected by the micro-TPC.
We also showed that more than 95$\%$ gamma-rays of the 511 keV 
were discriminated, 
while the efficiency to the neutrons of the same energy range 
was retained at $\sim$100$\%$.
Simulation studies showed that 
the total track length of the proton and the triton emitted from the 
$\rm {}^{3}He$(n,p(573 keV))$\rm {}^{3}H(191 keV)$ reaction
is about 1.2 cm, and that both particles 
have sufficient energy losses ($\rm > 200 keV/cm$) 
in 1 atm Ar+$\rm C_{2}H_{6}(10\%)$+${}^{3}$He($< 1\%$). 
These values suit the current performance of the micro-TPC, and 
we conclude that a time-resolved neutron PSD with 
a spatial resolution of sub-millimeters
shall be developed as an application of the micro-TPC.

\section*{Acknowledgment}
This work is supported by a Grant-in-Aid in Scientific Research of the Japan Ministry of Education, Culture, Science, Sports, Technology; ``Ground Research Announcement for Space Utilization'' promoted by Japan Space Forum; the joint research program with the high energy accelerator research organization(KEK); and the contract research program of the Japan Atomic Energy Research Institute(JAERI).


\newpage
\pagestyle{empty}

\newpage

\newpage

\begin{thebibliography}{1}

\bibitem{TPC_PSD:Kubo}
H.~Kubo, et al. proceedings of the 6th International Conference on Position Sensitive Detectors (PSD6), Leicester England, September 9-13, 2002, to appear in Nucl. Instr. Meth. A. 
\bibitem{IEEE:Miuchi}
K.~Miuchi, et al., IEEE Trans. Nucl. Sci., 50 (2003) 825.

\bibitem{netron_summary:Eijk}
Carel W. E. van Eijk, Nucl. Instr. Meth. A477 (2002) 383.


\bibitem{uPIC:Ochi}
A. Ochi, et al., Nucl. Instr. Meth. A 471(2001)264, A. Ochi, et al., Nucl. Instr. Meth. A478(2002)196, T.~Nagayoshi, et al.,  proceedings of the 6th International Conference on Position Sensitive Detectors (PSD6), Leicester England, September 9-13, 2002, to appear in  Nucl. Instr. Meth. A.

\bibitem{Maxwell}
Maxwell 3D Field Simulator, Ansoft Co.

\bibitem{Garfield}
Garfiled (ver7.03), R. Veenhof, Nucl. Instr. Meth. A419 (1998) 726.



\bibitem{ASD:Sasaki}
O. Sasaki and M. Yoshida, IEEE Trans. Nucl. Sci., 46 (1999).

\bibitem{Knoll}
G F. Knoll ''Radiation Detection and Measurement'', Third Edition, John Wiley \& Sons, Inc.





\bibitem{CCD-GEM:Fraga}
F. A. F. Fraga, et al., Nucl. Instr. Meth. A478 (2002) 357.


\bibitem{geant4}
Geant4, http://geant4.web.com.ch/geant4

\bibitem{ref:GEM_Fraga}
M.M.F.R. Fraga, et al., Nucl. Instr. Meth. A504 (2003) 88.

\bibitem{DRIFT_NIM}
D.P. Snowden-Ifft, T. Ohnuki, E.S. Rykoff, C.J. Martoff, Nucl. Instr. Meth. A498 (2003) 155.


\end{thebibliography}
\end{document}